\documentclass[preprint,nofootinbib,showpacs,aps,prc]{revtex4}

\usepackage{graphicx}
\usepackage{amstext}
\usepackage{amssymb}

\begin{document}

\title{Emission source functions in heavy ion collisions}

\author{V.M. Shapoval$^a$}
\author{Yu.M. Sinyukov$^a$}
\author{Iu.A. Karpenko$^{a,b}$}
\affiliation{$(a)$ Bogolyubov Institute for Theoretical Physics, Metrolohichna str. 14b, 03680 Kiev, Ukraine \\ $(b)$ Frankfurt~Institute~for~Advanced~Studies, Ruth-Moufang-Str.~1, 60438 Frankfurt am Main, Germany}

\begin{abstract}
The three-dimensional pion and kaon emission source functions are extracted from the HKM model simulations 
of the central Au+Au collisions at the top RHIC energy $\sqrt{s_{NN}}=200$~GeV. 
The model describes well the experimental data, previously obtained by the PHENIX and STAR collaborations using the imaging technique.
In particular, the HKM reproduces the non-Gaussian heavy tails of the source function in the pair transverse momentum (\textit{out}) and beam (\textit{long}) directions, observed in the pion case and practically absent for kaons. 
The role of the rescatterings and long-lived resonances decays in forming of the mentioned long range tails is investigated. 
The particle rescatterings contribution to the \textit{out} tail seems to be dominating. 
The model calculations also show the substantial relative emission times between pions (with mean value 14.5~fm/c
in LCMS), including those coming from resonance decays and rescatterings.
The prediction is made for the source functions in the LHC Pb+Pb collisions at $\sqrt{s_{NN}}=2.76$~TeV,
which are still not extracted from the measured correlation functions. 
\end{abstract}

\maketitle


Corresponding author: {\small \textit{Yu.M. Sinyukov, Bogolyubov Institute for Theoretical Physics, Kiev 03680, Metrolohichna 14b, Ukraine. E-mail: sinyukov@bitp.kiev.ua}}

\section{Introduction}
\label{part1}

The common method for accessing the spatio-temporal characteristics of the expanding superdense systems 
formed in the ultrarelativistic heavy-ion collisions is the correlation interferometry (or correlation femtoscopy) 
technique \cite{Gold,Kopylov,Coc}. It is based on the connection between the size and shape of the region, 
where the particles are produced (from the one side), and the form of the corresponding two-particle relative momentum distribution (from another side).
Physically this connection originates from the quantum statistics effect of symmetrization (antisymmetrization) 
of the two-particle wavefunction, leading to the enchancement (suppression) of the production of particles with close momenta.
The femtoscopic analysis utilizes the experimentally measured two-particle momentum correlation function, which is constructed as the ratio of the actual particle pairs distribution over the relative momentum (where pairs are formed by the particles from the same event) to the analogous distribution of the pairs of particles from mixed events. This correlation function is typically fitted with a certain analytical expression. 
In the most cases its quantum statistical component \footnote{Usually one supposes that contributions of the quantum statistical, the final state interaction and the non-femtoscopic correlations to the full correlation function can be separately accounted for in the fit~\cite{Factor1,Factor2}.} is supposed to be Gaussian $1+\lambda e^{-\sum_i q^2_i R^2_i}$, but in general 
case it depends on the researcher's assumptions about the emission function, thus being model-dependent. The correlation function fit gives one the interferometry radii $R_i$, commonly interpreted as the $i$-direction system's homogeneity lengths~\cite{Sin,AkkSin}.
It allows one to estimate the important characteristics of the dynamics of heavy ion collisions, such as  lifetime of the fireball created, gradients of the collective velocities, duration of emission, etc. 

In some sense a complementary method of obtaining the information about the space-time structure of the system from 
the correlation measurements, is known as the source imaging~\cite{BrownDan1,Stab,BrownDan2}. It is based on the extraction of the \textit{source function} -- the time-integrated distribution of the relative distance between particle radiation points in the pair rest frame (PRF) -- from the measured correlation function. In contrast to the standard approach, allowing to determine only the interferometry radii which are interpreted then in the previously assumed model, the source imaging reveals the actual non-Gaussian source function, being in this sense model-independent. Then, once the source function is obtained, one can readily fit it with different model expressions and extract the corresponding parameters. For example, fitting it with a Gaussian $\propto e^{-\sum_i r_i^2/(2 R_i^2)}$ one will find the $R_i$ values, which will be the same as in the standard femtoscopy method with the Gaussian fit for correlation function. But having the source function, apart from the Gaussian radii, one will also know the detailed source structure, which can likely deviate from the Gaussian distribution, having more complicated shape. This fact can be caused by different reasons, such as collision geometry, long-lived resonances contribution \cite{Sul,Nick,Csorgo}, space-momentum correlations due to either collective motion \cite{Pratt,Panit} or string fragmentation \cite{Anders}, etc. The source imaging in combination with collision model simulations
gives the possibility to study the influence of these effects on the source form. Another advantage of the imaging technique is that, in contrast to the correlation function, the source function reflects the properties of the emission region itself, refined from the FSI and QS effects.

In the paper \cite{Phenix} the authors investigate the source breakup dynamics
in the Au+Au collisions at $\sqrt{s_{NN}}=200$~GeV, analyzing two-pion source functions extracted from the 
experimental data. They observed a specific power-law tails in the pair-momentum and beam direction source
function projections, interpreted as the evidence for the noticeable emission duration time and long time delays
between emissions from different points of the source, in particular because of long-lived resonances decays. 
The kaon source function, obtained in the STAR collaboration Au+Au collision experiment at the similar energy \cite{Star} does not contain an analogous heavy tail, that probably indicates a less role of resonance's halo in the kaon emission and its almost instantaneous nature.

The tails observed in the pion source function can be interesting also in view of the activity
devoted to the search of the phase transition between the QGP and the hadron gas, which
could be expected to take place during the evolution of systems produced in the ultrarelativistic nuclear collisions \cite{Lacey}. The idea of such studies is that due to the soft equation of state (leading to the speed of sound $c_s = \left(\frac{\partial p}{\partial \epsilon}\right)_s^{1/2}$ close to zero, $c_s \approx 0$), which system has at the $1^{st}$ order phase transition, its expansion should slow down and the lifetime should increase. 
The source function's non-Gaussian tails can be considered as the possible signal of such prolonged system's lifetime.

To detail the interpretation of the extracted experimental data, one usually compares them
with the results of calculations in different event generators, such as Therminator \cite{Therm}
or Hadronic Rescattering Calculations~\cite{HRC} (HRC).
Based on such comparisons, one can see certain peculiarities of the system evolution. Another way of using the model source functions 
is including it into the correlation function fitting expressions, which account for the FSI effects,
such as Lednick\'y and Lyuboshitz analytical model \cite{Led}. Such models are utilized to fit the experimental data and extract the 
parameters characterizing the interaction between particles, such as scattering lengths, effective radii, etc.
This method allows one to find out the characteristics of, e.~g. strong interaction between particles of exotic species, that can hardly be measured in ordinary scattering experiments, but are accessible
for the FSI correlation technique in the relativistic heavy-ion collisions (see, e.g. \cite{fsi}).
The analytical approximations to the correlation function depend on both the source and the interaction characteristics,
that complicates the interpretation of the experimental data, increasing the number of free parameters in the fitting expression.
Calculating the parameters, that describe the source, separately in the event generator simulations could
facilitate and improve the reliability of the interaction analysis.

However, each of the mentioned models lacks completeness in the description of the matter evolution process
and involves a set of substantial simplifications. This complicates the analysis of the physical 
reasons of observed effects and most probably leads to missing of some features of the explored phenomena.
Say, in the most Therminator calculations the simple Blast-Wave \cite{BW} parametrization for the freeze-out hypersurface and flow is used, particle rescatterings are not implemented. The HRC includes rescatterings treatment, but accounts only for 8 sorts of resonances decays
and assumes kinematic evolution of particles just from the time $t=0$~fm/c, at which the simple parametrization for the initial particle momenta and coordinates distributions is exploited. 

In this paper we present the analysis of the source functions calculated in the hydrokinetic model (HKM)~\cite{HKM,Boltz,Kaon}, 
exactly, in its hybrid version (hHKM). The latter includes pure hydrodynamic stage, passing on to hydrokinetic one, describing the gradual liberation of particles from the expanding fluid, which is then switched on a space-like hypersurface to the UrQMD hadronic cascade. Thus, the model provides a realistic description of the full process of evolution of the matter produced in the relativistic nuclear collisions and is known to successfully describe a wide class of various observables \cite{Uniform}. 
The simulations were performed to describe the results for 20\% most central Au+Au collisions at $\sqrt{s_{NN}}=200$~GeV at RHIC and to make a prediction for 5\% most central LHC Pb+Pb collisions at $\sqrt{s_{NN}}=2.76$~TeV. 

In the section \ref{part2} we briefly discuss the physical meaning of the emission source function
and describe the idea of the source imaging technique. In the section \ref{part3} we show the results of our
calculations, compare it with experimental data and give our interpretation. In the section \ref{part4} we summarize our results and make concluding remarks.

\section{Source function and source imaging method}
\label{part2}

As follows from the previous section, the emission source function is an important object used in the analysis of the space-time structure of nuclear collisions.
In theoretical study based on the computer simulations one can easily extract the simulated source function directly from the event generator output,
as we do in the present article.
However, in the experiment one needs to utilize the source imaging technique to extract the source function from the measured correlation function $C(p,q)$.
To make the relation of the source function to the experimental observables more clear, 
in this section we remind to the reader the definition of the source function and the basic ideas of source imaging method.

Analyzing experimental data, the researcher aims to extract out of it all possible information about the explored object.
Studying particle emission in the relativistic nucleus-nucleus collision one would like to know
the emission function $g(x,p)=\frac{d^7N}{d^4x d^3p}$ -- the distribution of the emitted particles over the space-time coordinates and momentum components, that would provide exhaustive information about the analyzed process.
However, in practice it turns out, that the maximum possible knowledge about this process, obtainable in typical experiments, where the single-, two- and even many-particle momentum spectra are measured, 
is limited, so that $g(x,p)$ and even less informative Wigner function~$f_W(x,p)=\frac{d^6N}{d^3x d^3p}$ cannot be reconstructed in a model-independent way \cite{Emiss}.  

To see this, one can write the following expression for the Wigner phase-space density function~$f_W(x,p)$ \cite{Groot}
\begin{equation}
\label{Wig}
f_W(x,p) = (2 \pi)^{-3} \int d^4q \delta(q \cdot p) e^{-iqx} \langle a^{\dagger}_{p-q/2} a_{p+q/2} \rangle,
\end{equation}
justified for the case of weakly interacting particles.
Here $a^{\dagger}_k$, $a_k$ are the creation and annihilation operators of particles with momentum $k$,
$q=p_1-p_2$, $p=\frac{p_1+p_2}{2}$, and $p_1$, $p_2$ are the particle momenta.
The delta function $\delta(q \cdot p)$ under the integral sign corresponds to the assumed mass-shell constraint, and the brackets $\langle...\rangle$ mean averaging over the density matrix associated with the space-like hypersurfaces, at which particles become almost free (for sudden freeze-out the thermal density matrix at the freeze-out hypersurface can be used).

One can see, that to restore the Wigner function, one should have the possibility to extract 
the quantity $\langle a^{\dagger}_{p-q/2} a_{p+q/2} \rangle$. 
But can it be done, basing on the common experimental data?
Typically in the experiment one measures the single-particle $W(p)$ and the two-particle $W(p_1,p_2)$ momentum spectra 
to construct the two-particle correlation function
\begin{equation}
\label{cf0}
C(p_1,p_2) = \frac{W(p_1,p_2)}{W(p_1) W(p_2)}.
\end{equation}
The spectra can be expressed through $a^{\dagger}_p$ and $a_p$ 
\begin{eqnarray}\label{spectra}
 W(p) = E \frac{d^3N}{d^3p} & = & \langle a^{\dagger}_p a_p \rangle , \nonumber\\
 W(p_1, p_2) = E_1 E_2 \frac{d^6N}{d^3p_1 d^3p_2}& =& \langle a^{\dagger}_{p_1} a^{\dagger}_{p_2} a_{p_2} a_{p_1}\rangle = \nonumber\\ 
 & =& W(p_1) W(p_2) + \left| \langle a^{\dagger}_{p_1}a_{p_2}\rangle\right|^2,
\end{eqnarray}
where it is supposed, that four-operator average $\langle a^{\dagger}_{p_1} a^{\dagger}_{p_2} a_{p_2} a_{p_1}\rangle$ can be decomposed into the sum of products of two-operator ones
\begin{equation}
\langle a^{\dagger}_{p_1} a^{\dagger}_{p_2} a_{p_2} a_{p_1}\rangle = \langle a^{\dagger}_{p_1} a_{p_1} \rangle \langle a^{\dagger}_{p_2} a_{p_2} \rangle +
\langle a^{\dagger}_{p_1} a_{p_2} \rangle \langle a^{\dagger}_{p_2} a_{p_1} \rangle.
\end{equation}
From the Eq.~(\ref{spectra}) it follows, that having measured single- and two-particle momentum spectra, 
one can determine only the absolute value of $\langle a^{\dagger}_{p-q/2} a_{p+q/2} \rangle$,
so it will be known only to the phase factor 
\footnote{It can be shown, that the same situation takes place even if one includes many-particle spectra 
into consideration.}. 
This information is surely insufficient to restore the Wigner function (\ref{Wig}).

The emission function $g(x,p)$ is connected with the Wigner function through 
the integral equation
\begin{equation}
f_W(x,p) = f_W(x_0,p)+\int_{t_0}^t dt' g(x',p), 
\end{equation}
so its reconstruction makes even more hard task. That is why the analysis of the space-time structure
of the emission process has to be performed in terms of another (less informative) characteristics. 
The source function $S(\textbf{r}^{*})$ is one of them. It is usually defined as time-integrated pair separation distribution in the pair rest frame. 

In the experiment $S(\textbf{r}^{*})$ is restored from the measured correlation function $C(p,q)$. 
To figure out the connection between $S(\textbf{r}^{*})$ and $C(p,q)$, one can express both of them
via the emission function $g(x,p)$. If the smoothness approximation is assumed, 
the spectra (\ref{spectra}) can be written as
\begin{eqnarray}
\label{spectra2}
W(p) &=& \int d^4x g(x,p), \nonumber \\
W(p_1, p_2) &\approx & \int d^4x_1 d^4x_2 g_1(x_1,p_1) g_2(x_2,p_2) \left|\psi(\tilde{q},r)\right|^2,
\end{eqnarray}
where $\psi(\tilde{q},r)$ is a reduced Bethe-Salpeter amplitude corresponding to the relative motion of 
the particles making the pair with the generalized relative momentum $\tilde{q}=q-p(q p)/p^2$ and separation $r$. 
For the case of identical particles it should be replaced by the (anti)symmetrized amplitude, 
$\psi(\tilde{q},r) \rightarrow \left[\psi(q,r)\pm\psi(-q,r)\right]/\sqrt{2}$.
Here one supposes also that two-particle emission function is defined only by two-particle interaction,
whereas other effects, such as many-body interactions or event-wide correlations are neglected.

After substitution of (\ref{spectra2}) into (\ref{cf0}) the correlation function reads as
\begin{equation}\label{cf}
C(p,q) = 1 + \frac{\int d^4x_1 d^4x_2 g_1(x_1,p_1) g_2(x_2,p_2) 
\left(\left|\psi(\tilde{q},r)\right|^2-1\right)}{\int d^4x_1 g_1(x_1,p_1) \int d^4x_2 g_2(x_2,p_2)}.
\end{equation}
Then one can introduce the relative distance distribution function $s(r,p_1,p_2)$ as the convolution of normalized emission functions
\begin{eqnarray}
\label{rdd}
s(r,p_1,p_2) &=& \frac{\int d^4R g_1(R+r/2,p_1) g_2(R-r/2,p_2)}{\int d^4R g_1(R,p_1) \int d^4R g_2(R,p_2)} = \\ \nonumber
&=& \int d^4R \tilde{g_1}(R+r/2,p_1) \tilde{g_2}(R-r/2,p_2),
\end{eqnarray}
where tilde denotes the normalized emission function, $R = \frac{x_1 + x_2}{2}$, and $r = x_1 - x_2$. 

The femtoscopic correlations take place mainly between the particles with small relative particle velocities, $v_1 \approx v_2$. 
At this approximation for a pair with total momentum $P=p_1+p_2$ one has \cite{Led} $p_1 \approx \frac{m_1}{m_1+m_2} P$ and $p_2 \approx \frac{m_2}{m_1+m_2} P$ in (\ref{rdd}), 
so that $s(r,p_1,p_2)$ does not depend on $q$.
Assuming also the on-shell approximation, $P^0=\sqrt{(m_1+m_2)^2+\textbf{P}^2}$, one obtains
\begin{equation}
s(r,\textbf{P})=\int d^4R \tilde{g_1}(R+r/2,\frac{m_1}{m_1+m_2} \textbf{P}) \tilde{g_2}(R-r/2,\frac{m_2}{m_1+m_2} \textbf{P}).
\end{equation}
In the pair rest frame, where $\tilde{q} = \{0,\textbf{q}^{*}\}$, $\textbf{P}^*=0$, supposing the equal-time approximation to be justified, $t^{*} = t^{*}_1-t^{*}_2 = 0$ in the argument of $\psi(\tilde{q},r)$, 
one can substitute the Bethe-Salpeter amplitude in~(\ref{cf}) by the stationary solution of the scattering problem $\psi(\textbf{q}^{*},\textbf{r}^{*})$. 
Such substitution is valid provided that the condition \cite{Led} $|t^{*}| \ll m_{2,1} r^{*2}$ for $\mathrm{sign}(t^*)=\pm 1$ respectively is fulfilled. This is usually true for heavy particles such as kaons or protons. 
For pions produced in typical nuclear collisions,
the equal time approximation leads to a slight overestimation ($<5\%$) of the strong FSI effect and
it does not influence the leading zero-distance ($r^* \ll |a|$, $a$ is the pair Bohr radius) effect 
of the Coulomb FSI \cite{Led,Ledn2}.
Then one can connect the correlation function with the time-integrated $s(r^{*},\textbf{P}^{*})$, obtaining the so-called Koonin equation \cite{Fermi,Koonin,Gyulassy,Led,Ledn1,Ledn2,Ledn3}
\begin{eqnarray}\label{koon}
R(\textbf{q}^{*},\textbf{0})=C(\textbf{q}^{*},\textbf{0})-1=\int d^3r^{*} \int dt^* s(r^{*},\textbf{0}) (|\psi(\textbf{r}^{*},\textbf{q}^{*})|^2-1) = \int d^3r^{*} S(\textbf{r}^{*}) K(\textbf{r}^{*},\textbf{q}^{*}).
\end{eqnarray}
where the function $K(\textbf{r}^{*},\textbf{q}^{*})$ represents the kernel of the integral transform and $S(\textbf{r}^{*})=\int dt^* s(r^{*},\textbf{0})$ is the source function. 
The latter is interpreted as the probability density of emission of two particles at the relative distance $\textbf{r*}$ in their rest frame. 
Thus, it is normalized to unity
\begin{equation}
\int d^3 r^{*} S(\textbf{r*}) = 1. 
\end{equation}

Extraction of $S(\textbf{r*})$ from the correlation function (\ref{cf}) requires inverting the integral relation~(\ref{koon}). This operation is the central problem of imaging. It appears that computationally it is easier to reduce this three-dimensional problem to the series of one-dimensional ones. Such simplification can be achieved using the expansion of $R(\textbf{q*})$, $S(\textbf{r}^{*})$ and $K(\textbf{r}^{*},\textbf{q}^{*})$ in terms of spherical $Y_{lm} (\Omega)$ or Cartesian $A_{\textbf{l}}(\Omega)$ harmonics
\begin{equation}
Y_{lm} (\Omega) = \sqrt{\frac{2l+1}{4\pi} \frac{(l-m)!}{(l+m)!}} P^m_l(\cos \theta) e^{im \phi},
\end{equation}
where $P^m_l(\cos \theta)$ are associated Legendre polynomials, $l=0,1,2, \dots$ and $m=-l,\dots,l$;
\begin{eqnarray}
A_{\textbf{l}}(\Omega) = \sum_{m_i=0}^{l_i/2} \left(-\frac{1}{2}\right)^m \frac{(2l-2m-1)!!}{(2l-1)!!} \frac{l_x!}{(l_x-2m_x)!m_x!} \times \nonumber\\
\frac{l_y!}{(l_y-2m_y)!m_y!} \frac{l_z!}{(l_z-2m_z)!m_z!} n_x^{l_x-2m_x} n_y^{l_y-2m_y} n_z^{l_z-2m_z}.
\end{eqnarray}
Here $\textbf{n} = \{n_x,n_y,n_z\}$ is a unit vector in the $\Omega$ direction, $l_x+l_y+l_z = l$, $m_x+m_y+m_z = m$, $(-1)!!=1$.
Cartesian harmonics $A_{\textbf{l}}(\Omega)$ are linear combinations of spherical harmonics $Y_{lm} (\Omega)$ corresponding to one $l$ and different $m$.

The functions decomposition in, for example, spherical harmonics looks as
\begin{eqnarray}
\label{harm}
R(\textbf{q}^{*})&=&\sqrt{4\pi} \sum_{lm} R^*_{lm}(q^{*}) Y_{lm} (\Omega_{\textbf{q}^{*}}), \nonumber \\
S(\textbf{r}^{*})&=&\sqrt{4\pi} \sum_{lm} S^*_{lm}(r^{*}) Y_{lm} (\Omega_{\textbf{r}^{*}}), \nonumber \\
K(\textbf{q}^{*},\textbf{r}^{*}) &=& 4\pi \sum_{lm} K_l(q^{*},r^{*}) Y_{lm}(\Omega_{\textbf{q}^{*}}) Y^*_{lm}(\Omega_{\textbf{r}^{*}}),
\end{eqnarray}
where $R_{lm}(q^{*})=\frac{1}{\sqrt{4\pi}} \int d\Omega_{\textbf{q}^{*}} Y_{lm} (\Omega_{\textbf{q}^{*}}) R(\textbf{q}^{*})$ 
and $S_{lm}(r^{*})=\frac{1}{\sqrt{4\pi}} \int d\Omega_{\textbf{r}^{*}} Y_{lm} (\Omega_{\textbf{r}^{*}}) S(\textbf{r}^{*})$
are called spherical \textit{correlation} and \textit{source} moments respectively. 
Substitution of the obtained expansions (\ref{harm}) into the equation (\ref{koon}) 
gives one the series of one-dimensional integral equations with respect to source function 
spherical moments $S_{lm}(r^{*})$
\begin{equation}\label{onedim}
R_{lm}(q^{*}) = 4\pi \int dr^{*} r^{*2} S_{lm}(r^{*}) K_l(q^{*},r^{*}).
\end{equation}

Apart from simplifying imaging calculations, the decomposition of correlation function in harmonics,
in contrast to its one-dimensional projections, represents
the complete information about three-dimensional correlation structure and provides focused insight to the specific physical properties of the emission process (see, e.g. \cite{harms}). 

For instance, $R_{00}$ moment represents angle-integrated correlation, being sensitive mainly to invariant radius $R_{inv}$. The moments corresponding to $l=1$ provide access to ``Lednicky offset'' \cite{Offset}, telling which sort of the particles was emitted earlier. They vanish in the case of identical particles due to symmetry. $R_{20}$ is sensitive to the ratio of transversal and longitudinal source sizes. The more it differs from zero, the stronger asymmetry between these sizes takes place. The moment $R_{22}$ corresponds to the outward to sideward system sizes ratio. As for the moments with $l=3$, they contain information about so-called "boomerang" triaxial deformation and also disappear for identical particles. The moments with $l \ge 4$ provide rather detailed information about the source and are not intensively studied at the moment.
Also, the harmonics decomposition is more directly connected with the source shape \cite{DanPratt} and simplifies the analysis of non-femtoscopic correlations \cite{Chaj}.

As for solving the system (\ref{onedim}), in the most simple cases, when the analyzed correlation function contains the information only about the quantum statistics correlations,  it can be performed analytically, using the inverse integral transform. In this case the kernel moments $K_l(q^*,r^*)$, containing spherical Bessel functions $j_l$, define the Fourier-Bessel transform, which can be easily inverted using the completeness relation for $j_l$
\begin{equation}
\label{invert}
S_{lm}(r^*)=4 \pi \int_0^\infty dq^* q^{*2} K_l^{-1}(q^*,r^*) R_{lm}(q^*),
\end{equation}
where $K_l^{-1}(q^*,r^*)$ is the inverse transform kernel.
But in general case, when the kernel $K(\textbf{q}^*,\textbf{r}^*)$ is more complicated, $K_l^{-1}(q^*,r^*)$ 
cannot be found in analytical form. Instead of finding it, one can discretize the Eq. (\ref{onedim})
and solve it numerically
\begin{equation}
R_{lm}(q_i) = 4\pi \sum_j \Delta r r_j^2 S_{lm}(r_j) K_l(q_i,r_j) = \sum_j K^l_{ij} S^{lm}_j 
\end{equation}
where $i=1, \dots, N_q$, $j=1,\dots,N_r$ with $N_q$ and $N_r$ being the numbers of discrete points where the values of $R(q^*)$ and $S(r^*)$ are specified. 
Since generally $N_q \neq N_r$, the obtained system of linear equations can be under- or overdetermined. 
So, one usually tries to find the values $S_j$ (for each $l,m$) using the method of $\chi^2$-minimization
\begin{equation}
\chi^2 = \sum_i \frac{(\sum_j K_{ij} S_j - R_i)^2}{\sigma_i^2}
\end{equation}
with $\sigma^2$ being the variance of observed correlation moments $R_{lm}$.
Equating the $\chi^2$ derivative with respect to $S_j$ to zero one obtains the following series of equations
\begin{equation}
\sum_{ij} \frac{1}{\sigma_i^2} (K_{ij} S_j - R_i) K_{ij} = 0.
\end{equation}
Its solution in the matrix form is \cite{BrownDan1}
\begin{equation}
S = (K^T K)^{-1} K^T R.
\end{equation}

However, the integral equation (\ref{onedim}) that has to be solved is a homogeneous Fredholm integral equation, which is actually an ill-posed problem for numerical solution because of the singular or ill-conditioned $K$ matrix. Singularity in the context of numerical calculations means that the matrix has one or more eigenvalues negligibly small as compared to the others. 
It leads to instability of the resulting solution, i.e. the small uncertainties in the $R_i$ values can cause big uncertainties in $S_j$. Thus, the solution will not be a smooth function and will be different depending on the used solving algorithm.
This problem is quite general and well-known for the whole inverse problem class. There are certain methods, which can be applied to increase the solution stability \cite{Stab}, but its successful application, resulting in unambiguous imaging problem solution, i.e. extracting the source function from the experimental data, is a non-trivial and complicated task.


Fortunately, to obtain the model source function from the event generator one does not need 
to utilize the imaging technique, since in this case the source function can be extracted from
the program output straightforwardly. The next section contains the results of source functions
calculation in the hydrokinetic model, its interpretation and comparison with the experiment.

\section{Results and discussion}
\label{part3} 

Hydrokinetic model \cite{HKM,Boltz,Kaon} simulates the evolution of matter formed in the relativistic heavy-ion collisions.
The full process is supposed to pass in two stages -- a continuous medium expansion, described in the ideal hydrodynamics approximation, which then goes over to gradual system decoupling, described in the hydrokinetic
approach. At the first stage matter is supposed to be in local chemical and thermal equilibrium. Here we use the lattice-QCD inspired equation of state for the quark-gluon phase, matched via the cross-over type transition with the hadron resonance gas, consisting of all 329 well-established hadron states made of u, d, s quarks.
As the system expands and cools down, it reaches the second stage, supposed to begin at the chemical freeze-out isotherm $T_{ch}=165$~MeV. At temperatures $T<T_{ch}$ system gradually loses both chemical and thermal equilibrium,
and the particles begin to continuously escape from the medium. 
In the hybrid model version (hHKM) the hydrokinetic description of the second stage is switched to the UrQMD hadron cascade on a space-like hypersurface, situated behind the hadronization one. Another option consists in direct switching to the cascade just from the hydrodynamics, at the hadronization hypersurface $T_{ch}=165$~MeV. 
We use this particular variant in the current analysis, relying on the result of \cite{Uniform}, where the comparison of one- and two-particle spectra, calculated at both types of matching hydro and cascade stages, showed a quite small
difference between them in the considered case of top RHIC and LHC energies.

At the switching hypersurface a set of particles is generated according to the corresponding distribution function
using either Cooper-Frye prescription \cite{Cooper} (for sudden switching from hydro to UrQMD) or the technique of Boltzmann equations in integral form \cite{Boltz,Kaon} (if hydrokinetics is involved). This set serves as input for UrQMD, which performs particle rescatterings and decays. The final model output is again a collection of particles, characterized by their momenta and the points of their last collision.

HKM showed itself to be successful in a simultaneous description of kaon and pion femtoscopy together with corresponding momentum spectra at top RHIC and LHC energies~\cite{Kaon,Uniform}. So we can expect the source 
functions extracted from the model also to be realistic and reliable. Here we present the pion and kaon source functions generated by HKM using parameters adjusted for description of the data from RHIC 20\% most central Au+Au collisions at $\sqrt{s}= 200$~GeV. Also the predictions concerning the LHC 5\% most central Pb+Pb collisions at $\sqrt{s}=2.76$~TeV are demonstrated.

We work in the central rapidity slice and assume longitudinal boost-invariance. Early thermalization at proper time $\tau = 0.1$ fm/c is supposed. In transverse plane we use Glauber MC initial energy density profile generated by GLISSANDO code \cite{glissando} and suppose small but non-zero initial transverse flow which is taken linear in transversal radius $r_T$ \cite{Uniform}: $y_T = \alpha \frac{r_T}{R^2(\phi)}$. Here $R(\phi)$ is a system's homogeneity length in $\phi$-direction, $\phi$ is an azimuthal angle. Such a small initial flow mimics the shear viscosity effects at the system evolution based on the perfect hydrodynamics
as well as the effects of event by event fluctuating hydro-solutions \cite{Uniform}. 
Maximal initial energy density $\epsilon_0$ is chosen to reproduce the experimental mean charged particle multiplicity. Thus, $\epsilon_0$ 
and the coefficient $\alpha$ are the only fitting parameters of the model. For the case of Au+Au collisions at $\sqrt{s}= 200$ GeV we take
the parameters from \cite{Uniform} corresponded to the best fit for the pion, kaon and proton spectra and pion interferometry data, 
$\epsilon_0 = 430$~GeV/fm$^3$ and $\alpha = 0.45$~fm (the maximal initial transverse velocity at the very periphery of the system is then 0.05).

The source functions are calculated from the event-generator output according to the formula
\begin{equation}
S(\textbf{r}^{*})=\frac{\sum_{i \ne j}\delta_{\Delta}(\textbf{r}^{*}-\textbf{r}^*_i+\textbf{r}^*_j)}{\sum_{i \ne j} 1}.
\end{equation}
Here $\textbf{r}^*_i$ and $\textbf{r}^*_j$ are the particles space positions, and $\textbf{r}^*$ is the particles separation
in the pair rest frame, $\delta_{\Delta}(x) = 1$ if $|x|<\Delta p/2$ and 0 otherwise, $\Delta p$ is the size of the histogram bin.

In the PHENIX experiment \cite{Phenix} the measured correlation function is decomposed into Cartesian correlation moments
\begin{equation}
R(\textbf{q}) = \sum_l \sum_{\alpha_1 \ldots \alpha_l}
   R^l_{\alpha_1 \ldots \alpha_l}(q) \,A^l_{\alpha_1 \ldots \alpha_l} (\Omega_{\textbf{q}}),
\label{eqn1}
\end{equation}
where $l=0,1,2,\ldots$, $\alpha_i=x, y \mbox{ or } z$, 
$A^l_{\alpha_1 \ldots \alpha_l}(\Omega_{\textbf{q}})$
are Cartesian harmonic basis elements, ($\Omega_{\textbf{q}}$ is the solid 
angle in $\textbf{q}$ space), and $R^l_{\alpha_1 \ldots \alpha_l}(q)$ are 
Cartesian correlation moments given by 
\begin{equation}
 R^l_{\alpha_1 \ldots \alpha_l}(q) = \frac{(2l+1)!!}{l!}
 \int \frac{d \Omega_{\textbf{q}}}{4\pi} A^l_{\alpha_1 \ldots \alpha_l} 
 (\Omega_{\textbf{q}}) \, R(\textbf{q}).
 \label{eqn2}
\end{equation}
The obtained correlation moments then are used as the input data for the source imaging method.
In \cite{Phenix} only the even moments up to order $l=6$ were utilized, 
whereas the odd moments were found to be consistent with zero,
as it was expected from symmetry considerations, and the moments of higher order were found to be negligible.

\begin{figure}
\begin{center}
\includegraphics[scale=0.82]{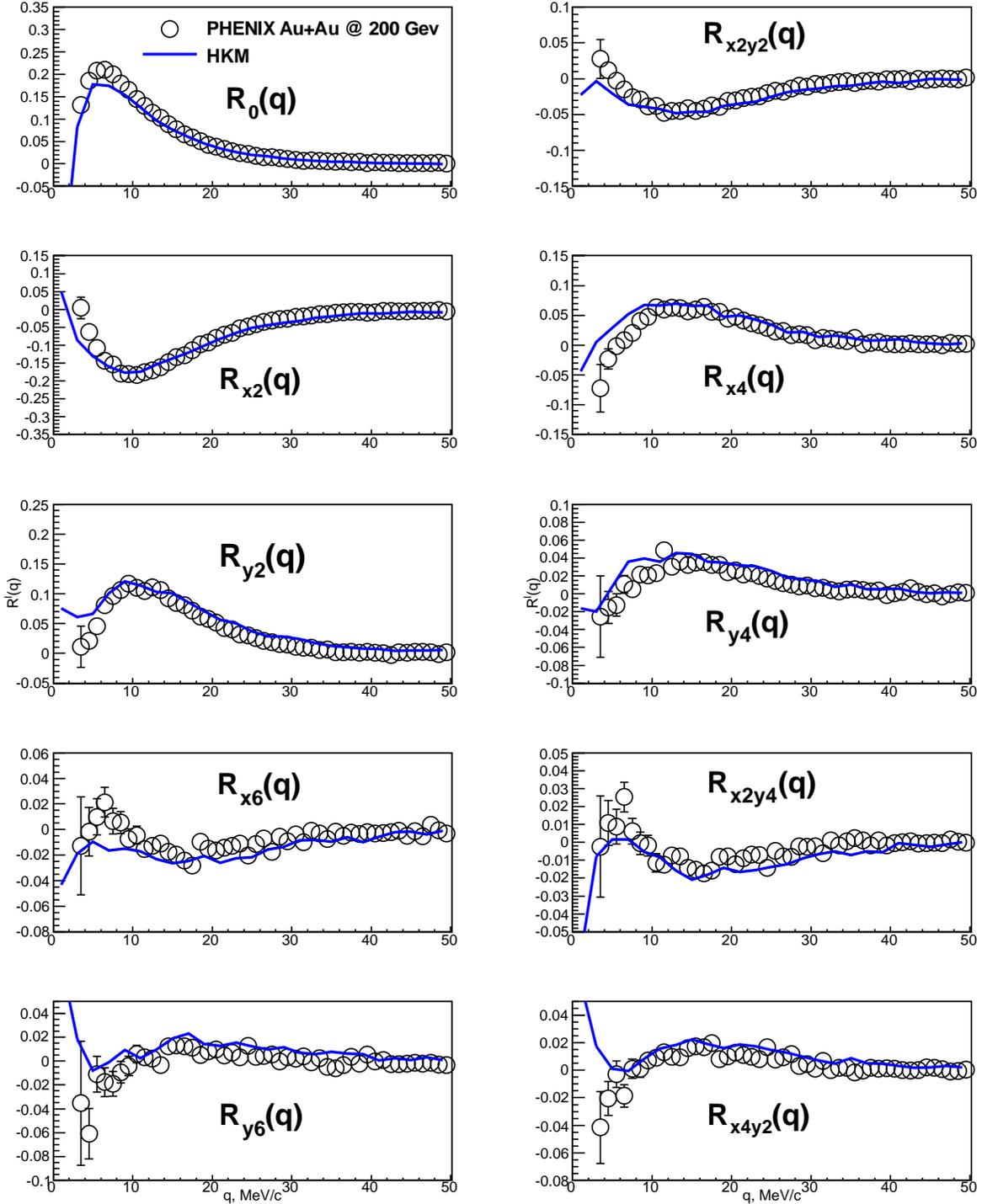}
\end{center}
\caption{\label{fig1}
Two-pion correlation moments $R_{xl_xyl_y}(q)$ ($q$ is in MeV/c) obtained from the HKM model (blue lines) and from the data measured by PHENIX (open circles), 
$0.2<p_T<0.36$~GeV/c, $|y|<0.35$, $c=0-20$\%.}
\end{figure}

To demonstrate the model reliability in describing the experimental data we compare in Fig. \ref{fig1} the
simulated two-pion Cartesian correlation moments with the experimental ones. 
Among all the moments used in the correlation function decomposition only 10 moments presented in the figure
are independent, and the rest can be expressed through these 10 ones.
Here $R_{xl_xyl_y}$ denotes the Cartesian correlation moment corresponding to $l=l_x+l_y$, $\alpha_1 = \dots = \alpha_{l_x} = x$, $\alpha_{l_x+1} = \dots = \alpha_{l} = y$ .
As one can see, the HKM calculations result is in a good agreement with the experiment.

\begin{figure}
\begin{center}
\includegraphics[scale=0.82]{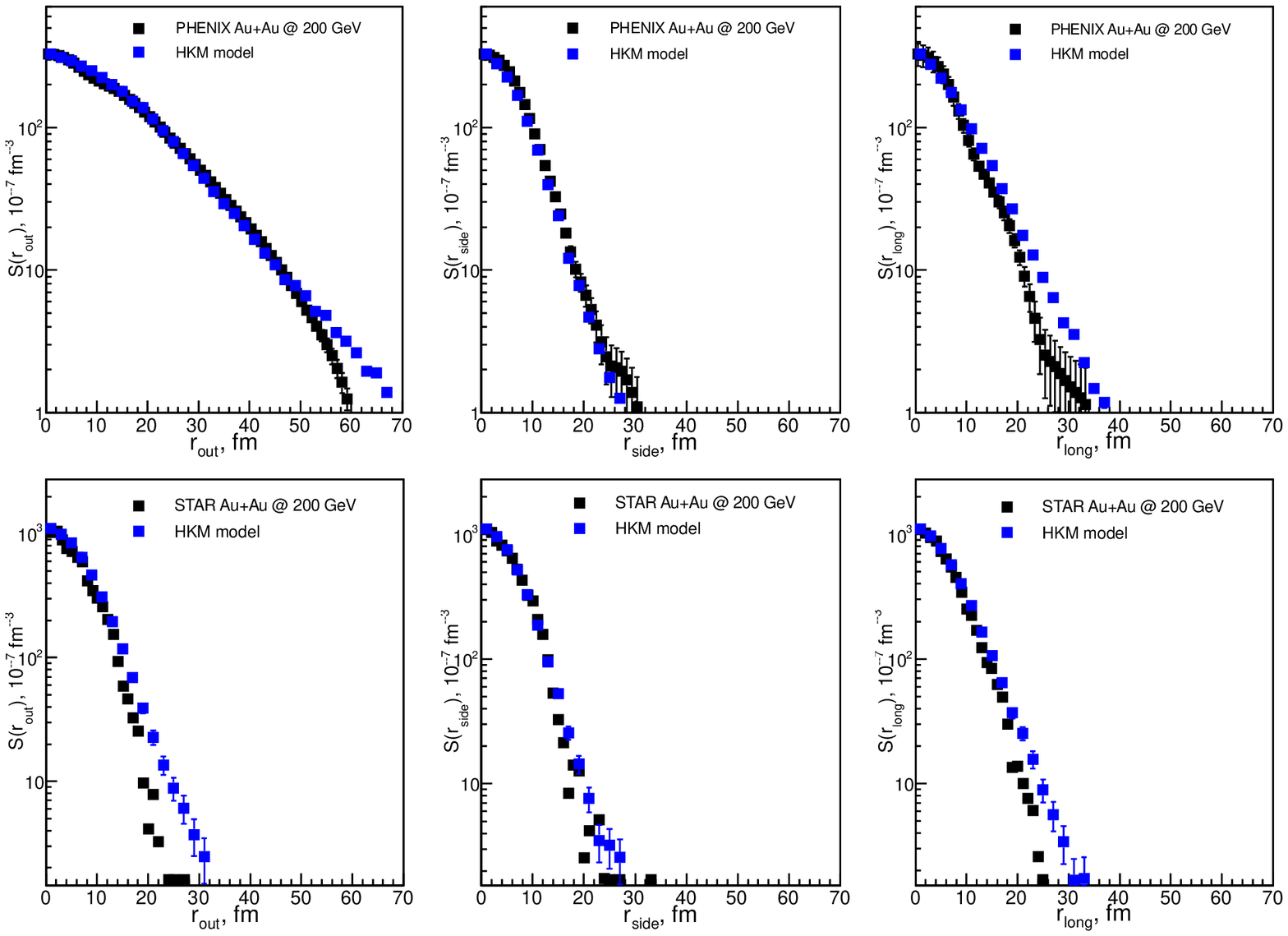}
\end{center}
\caption{\label{fig2}
Pion (top) and kaon (bottom) source function projections extracted from the HKM model (blue markers)
compared with the ones obtained by PHENIX and STAR from the experimental correlation data using imaging 
procedure (black markers), $0.2<p_T<0.36$~GeV/c, $|y|<0.35$ for pions and $|y|<0.5$ for kaons, $c=0-20$\%.}
\end{figure}

The comparison of projections of the three-dimensional pion and kaon source functions calculated in HKM with the PHENIX and STAR experimental data can be found in Fig.~\ref{fig2}~
\footnote
{
The demonstrated HKM source functions (as well as the correlation moments in the Fig. \ref{fig1}) are scaled by a factor $\lambda_{exp}/\lambda_{\mathrm{HKM}}<1$ 
being the ratio of the experimental $\lambda_{exp}$ and the model $\lambda_{\mathrm{HKM}}$ correlation function suppression parameters, which define also the source function intercept.
In the present study we do not aim to analyze in detail the reasons of disagreement between both $\lambda$
values (the smaller $\lambda_{exp}$ value could be caused by, e.g. misidentification of certain fraction of particles in the experiment etc.).
Instead we would like to focus on exploring the space-time extent of the source, 
reflected in the \textit{shape} of the source function, which is reproduced well in our simulations as one can see.
}.
Here the Bertsch-Pratt coordinate system is used, where the \textit{out} axis is directed along the pair total momentum, the \textit{long} direction coincides with the beam axis, and the \textit{side} axis is perpendicular to the latter two ones.

We can see that the source functions are reasonably good reproduced by HKM both for pions and kaons, 
including the non-gaussian tails in \textit{out} and \textit{long} directions for pion case. 
The range of $S(r_{long})$ projection reflects the source lifetime, and $S(r_{out})$ is 
affected by several factors. 
Since \textit{out} direction corresponds to the pair momentum in longitudinally co-moving system (LCMS), 
the tail in $S(r_{out})$ can be partially explained by Lorentz dilation at boost from LCMS to pair center of mass frame (PCMS).
However, the estimates made in \cite{Phenix2} show that even at the maximal Lorentz dilation,
when $R_{out}$ value in PCMS is $\gamma$ times larger than the LCMS one, the observed long-range
source component cannot be explained only by such kinematic transformation.
Another possible explanation consists in associating the long-range tail with long 
delays between particle emission times due to halo of secondary particles from long-lived resonances
decays and due to hadron rescatterings. The latter may cause under certain conditions the so-called anomalous diffusion of particles, 
characterized by increasing in time mean free path \cite{AnDiff}. 

\begin{figure}
\begin{center}
\includegraphics[scale=0.82]{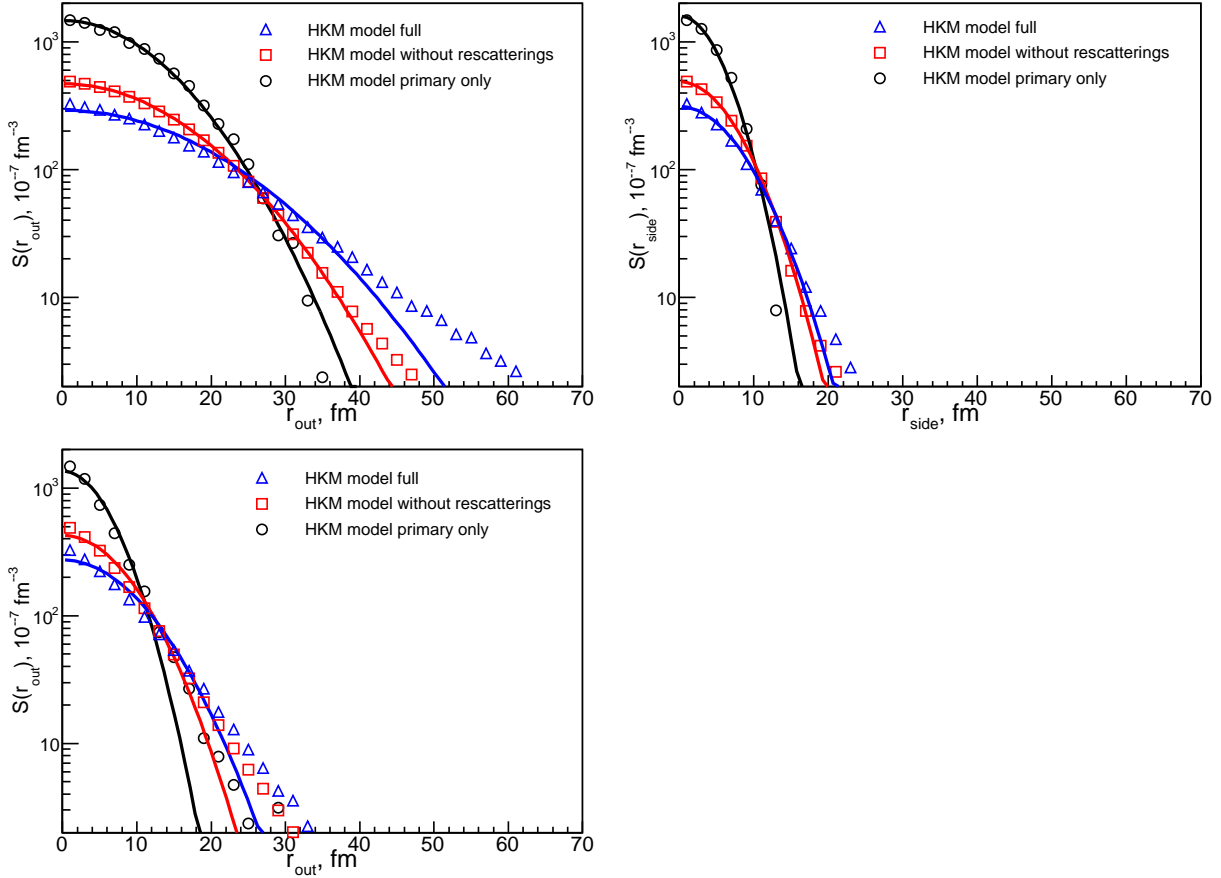}
\end{center}
\caption{
\label{fig3}
The HKM pion source functions constructed from the full model output (blue triangles), model output without
rescatterings (red rectangles) and the primary particles only (black circles), $0.2<p_T<0.36$~GeV/c, $|y|<0.35$.
Solid lines corresponds to the Gaussian fits to corresponding HKM results.}
\end{figure}

Let us investigate the contributions of secondary particles from the long-lived resonances decays and particle
rescatterings to the formation of observed heavy tail. In Fig.~\ref{fig3} we present three source functions,
constructed from different model outputs (open markers): the full one, the model output with rescatterings turned off, 
and the one including only primary particles, i.e. without both rescatterings and resonances decays. 
Along with the model data we also show the corresponding Gaussian fits (solid lines).
Considering the \textit{out}-direction profile, one can see that the full model output has a large non-Gaussian tail, 
while in the case of no rescatterings it is quite reduced. 
The primary particles source function is almost Gaussian in transverse direction.
It seems that the exclusion of rescatterings reduces the \textit{out} tail more significantly, than the exclusion of 
resonances decays contribution, so the rescatterings seem to play the main role in the tail formation.
Also one can conclude that the role of Lorentz boost to PCMS should be of minor importance, since the tail
is already practically absent for the source function built from primary particles in PCMS, so that
accounting for the Lorentz transformation can only change the Gaussian interferometry radius $R_{out}$,
but cannot additionally reduce the $S(r_{out})$ deviation from Gaussian form by any noticeable extent.

As for the interpreting the observed tails as the signals of the phase transition, the situation is not so easy.
The different criteria for detecting the transition occurrence were proposed, such as the Bertsch criterion 
$R_{out}/R_{side} \gg 1$ or large $R_{long}$ criterion. However, in practice there exist certain difficulties 
in its application. For example, if the matter expansion starts from the state of pure phase 
and then at some stage of the evolution the system undergoes a phase transition, the fast matter expansion will continue of inertia, 
and we will not find the transition signs in the measured data. To see these signs, one would have to create such initial conditions 
for matter evolution, that it would start from the transient state, where different phases exist simultaneously. 
Apart from this, the application of the Bertsch criterion is troubled by likely existence of positive $r_{out} - t$ 
correlation in the sectors of the freeze-out hypersurface, where the particles with considered momenta come from, 
that causes the negative sign of the space-time correlation contribution to $R_{out}$ and in such a way reduces the observed $R_{out}$ value \cite{RoRs1,RoRs2}
\begin{equation}
R^2_{out}(p)=\langle (\Delta r_{out} - v_{out}\Delta t)^2 \rangle_p=\langle \Delta r^2_{out}\rangle_p+
v^2_{out}\langle\Delta t^2\rangle_p - 2 v_{out}\langle \Delta r_{out} \Delta t \rangle_p.
\end{equation}
This means that the observed $R_{out}/R_{side}\approx 1$ ratio does not necessarily indicate the short emission duration. 
However, one can try to extract the mean emission duration as well as system's lifetime from the event generator calculations.

In \cite{Phenix} the experimental data for pions are presented together with results of simulations in Therminator event generator with Blast-Wave parametrization for the freeze-out hypersurface, $\tau = \tau_0 + a \rho$, where 
$\tau_0$ is the source proper breakup time, $\rho$ is transverse radial coordinate, $\rho \leq \rho_{max}$, and $a$ is a free parameter, describing space-time correlations.
The simulations give the best data description, when the resonances decays are turned on and particle emission
is supposed to take place from the family of hypersurfaces defined by different breakup proper times~$\tau_0'$, 
distributed according to exponential law, $\frac{dN}{d\tau_0'} = \frac{\Theta(\tau_0'-\tau_0)}{\Delta \tau} \exp\left(\frac{\tau_0'-\tau_0}{\Delta \tau}\right)$. 
The width $\Delta \tau$ is interpreted then as mean emission duration time in the rest frame 
\footnote
{
One can say that such parametrization efficiently accounts for particle rescatterings, 
which are not implemented in Therminator, but seem to strongly influence the measured source function. 
}.
The break-up proper time $\tau_0 \sim 9$~fm/c and small but non-zero proper emission duration in the rest frame $\Delta \tau \sim 2$~fm/c, 
at which simulations results are most close to the data, appear to be incompatible with the first order transition scenario \cite{Pratt2,Rischke}, but can point to cross over phase transition \cite{Lacey}.
Finite emission duration in the used parametrization (and, partly, the resonance decays contribution) leads to quite appreciable pion emission time differences in LCMS, with $\langle|\Delta t_{\mathrm{LCMS}}|\rangle \approx 12$~fm/c.

As for the hybrid HKM, the cross over phase transition is assumed in the model from the beginning
in the equation of state for the hydro stage. We use the isotherm $T_{ch}$ as the single freeze-out hypersurface, which in the case of top RHIC energies spreads in proper time up to $\tau_{max} \approx 7$~fm/c,
and the mean LCMS emission time difference for pions in our calculations is $\langle|\Delta t_{\mathrm{LCMS}}|\rangle=14.5$~fm/c including emission from resonances and after rescatterings, that is apparently in accordance with the results obtained within the Therminator parametrization.

\begin{figure}
\centering
\includegraphics[scale=0.82]{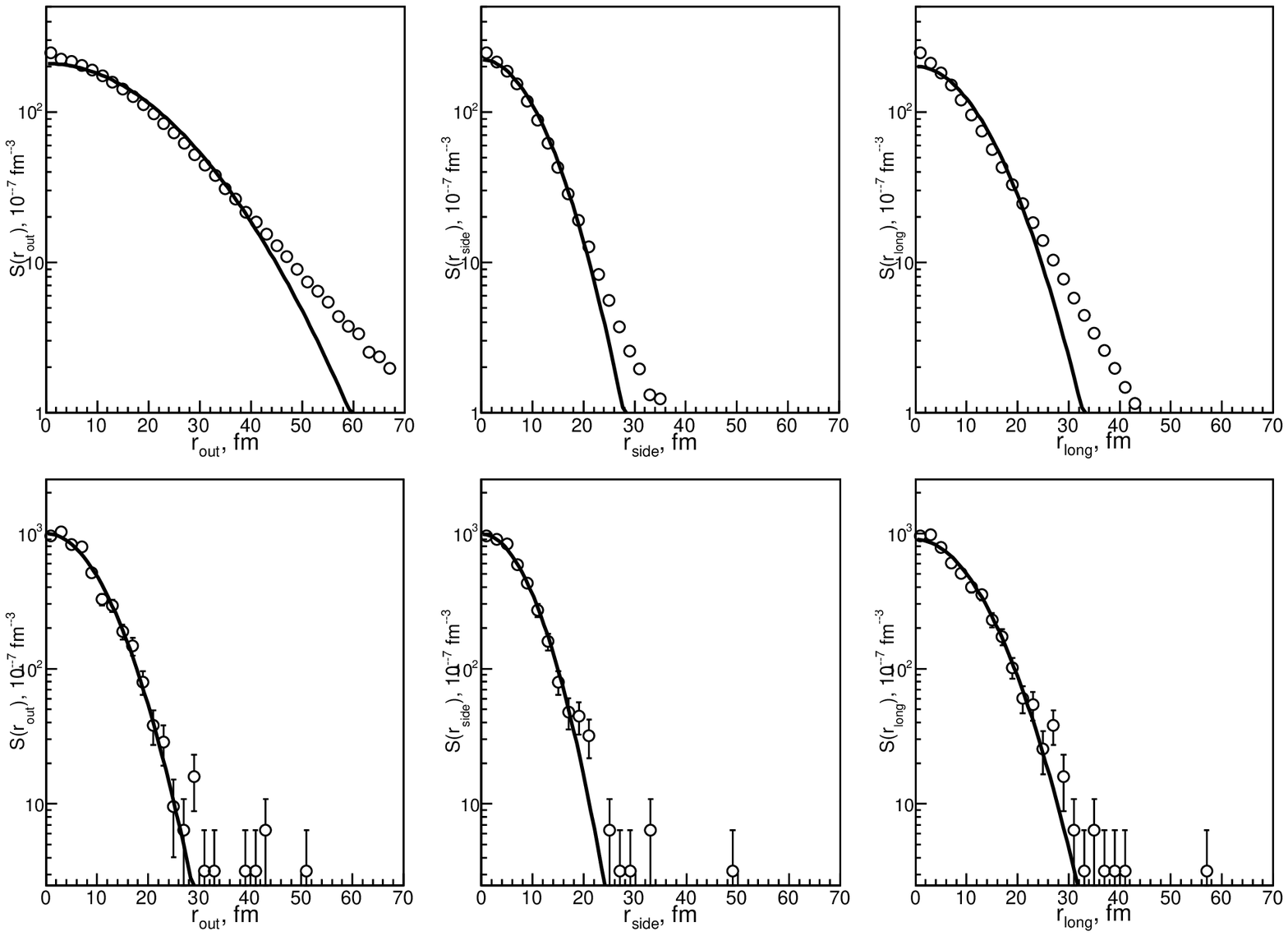}
\caption{\label{fig4}
Pion (top row) and kaon (bottom row) source function projections from HKM with parameters adjusted for Pb+Pb LHC collisions at $\sqrt{s} = $2.76 TeV, $c=0-5$\%, $0.2<p_T<0.36$~GeV/c, $|y|<0.5$. Solid lines represent the Gaussian fits to the corresponding HKM calculations, presented by circles. }
\end{figure}

Finally, in Fig. \ref{fig4} we show our predictions concerning source function 
for the case of 5\% most central LHC Pb+Pb collisions at the energy $\sqrt{s} = $2.76 TeV.
For these calculations we take the same $\alpha=0.45$~fm as for the RHIC case, and the maximal energy density at starting time $\tau=0.1$~fm/c is equal to $\epsilon = 1300$~GeV/fm$^3$.
We see that according to our calculations, the kaon source functions at LHC also have to be closer to the
Gaussian shape, whereas for the pion case, the source function should have a heavy tail.


\section{Conclusions}
\label{part4} 
In this paper we considered the analysis of heavy-ion collision space-time
structure in terms of emission source functions, or time-integrated particle 
emission points separation distribution related to the pair rest frames. 
Experimental source functions serve as model-independent characteristics
of the size and shape of the emission region, separated from the particle interaction effects. 
Restored from the measured correlation functions with a sophisticated
source imaging technique, experimental source functions can be compared 
with the model ones, directly extracted from the event generators of a particular model.

Here we presented the results of pion and kaon source function simulations within hydrokinetic model (HKM)
for the semi-central Au+Au collisions at top RHIC energy $\sqrt{s_{NN}}=200$~GeV together with the predictions
for the central LHC Pb+Pb collisions at $\sqrt{s_{NN}}=2.76$~TeV.
The results for RHIC are in good agreement with the experimental data.
The long-range power-law tail in the pair momentum direction, observed in pion source functions, 
can be explained by combined influence of secondary particles coming from the long-lived
resonances decays and particle rescatterings, whose contribution seems to be dominating.
The role of the kinematic transformation to the pair rest frame seems to be of minor importance.
The model calculations also point to substantial emission time differences for pion pairs 
that should take place in the experiment.

The behavior of the predicted source functions for the LHC case is qualitatively similar to the RHIC one,
including the presence of heavy tails for pions and almost Gaussian source shape for kaons.

\section{Acknowledgement}
\label{part5} 
The authors are grateful to M. Sumbera and P. Chung for fruitful discussions. 
The research was carried out within the scope of the EUREA:
European Ultra Relativistic Energies Agreement (European
Research Group: ``Heavy ions at ultrarelativistic energies'')
and is supported by the National Academy of Sciences of
Ukraine (Agreement F4-2013) and by the State Fund for
Fundamental Researches of Ukraine (Agreement F33/24-2013).
The work was supported by the Program of Fundamental
Research of the Department of Physics and Astronomy of the
NAS of Ukraine.


\newpage
\begin{thebibliography}{99}

\bibitem{Gold} G. Goldhaber, S. Goldhaber, W. Lee, A. Pais, Phys. Rev. {\bf 120}, 325 (1960).

\bibitem{Kopylov} G.I. Kopylov, M.I. Podgoretsky, Sov. J. Nucl. Phys. {\bf 15}, 219 (1972); {\bf 15}, 219 (1972),  
{\bf 18} 336, (1973); {\bf 19}, 215 (1974).

\bibitem{Coc} G. Cocconi, Phys. Lett.  {\bf B49,} 459 (1974).

\bibitem{Factor1} Yu.M. Sinyukov, R. Lednick\'y, S.V. Akkelin, J. Pluta, and B. Erazmus, Phys. Lett. B \textbf{432}, 248 (1998).

\bibitem{Factor2} Yu.~M.~Sinyukov, V.~M.~Shapoval, Phys. Rev. D {\bf 87}, 094024 (2013), arXiv:1209.1747.

\bibitem{Sin}Yu.M. Sinyukov, Nucl. Phys. A {\bf 566}, 589 (1994);
in \textit{Hot Hadronic Matter: Theory and Experiment}, edited by
J. Letessier, H.H. Gutbrod, and J. Rafelski (Plenum, New York, 1995), p. 309.

\bibitem{AkkSin} S.V. Akkelin, Yu.M. Sinyukov, Phys. Lett. B {\bf 356},  525 (1995).

\bibitem{BrownDan1} D.A. Brown and P. Danielewicz, Phys. Lett. B \textbf{398}, 252 (1997).

\bibitem{Stab} D.A. Brown and P. Danielewicz, Phys. Rev. C \textbf{57}, 2474 (1998).

\bibitem{BrownDan2} D.A. Brown and P. Danielewicz, Phys. Rev. C \textbf{64}, 14902 (2001).

\bibitem{Sul} J.~P. Sullivan \textit{et al.}, Phys. Rev. Lett. \textbf{70}, 3000 (1993).

\bibitem{Nick} S. Nickerson, T. Cs\"org\H{o}, and D. Kiang, Phys. Rev. C \textbf{57}, 3251, (1998).

\bibitem{Csorgo} T. Cs\"org\H{o}, B. Lorstad, J. Schmid-Sorensen, and A. Ster, Eur. Phys. J. C \textbf{9}, 275, (1999).

\bibitem{Pratt} S. Pratt, T. Cs\"org\H{o}, and T. Zim\'anyi, Phys. Rev. C \textbf{42}, 2646 (1990).

\bibitem{Panit} S.~Y. Panitkin and D.~A. Brown, Phys. Rev. C \textbf{61}, 021901(R) (2000).

\bibitem{Anders} B. Andersson and W. Hofmann, Phys. Lett. \textbf{169B}, 364 (1986).

\bibitem{Phenix} S. Afanasiev \textit{et al.} PHENIX Collaboration, Phys. Rev. Lett. 100, 232301 (2008). 

\bibitem{Star} P. Chung for The STAR Collaboration, arXiv:1012.5674 [nucl-ex]; \\
               L. Adamczyk \textit{et al.} STAR Collaboration, arXiv:1302.3168 [nucl-ex].

\bibitem{Lacey} Roy A. Lacey, Braz. J. Phys. \textbf{37}, 893 (2007); J. Phys. G \textbf{35}, 104139 (2008).

\bibitem{Therm} A. Kisiel, T. Taluc, W. Broniowski and W. Florkowski, Comput. Phys. Commun. \textbf{174}, 669 (2006).

\bibitem{HRC} T. J. Humanic, Nucl. Phys. A \textbf{715}, 641, (2003); Phys. Rev. C \textbf{73}, 054902 (2006).
             
\bibitem{Led} R. Lednick\'y, V. L. Lyuboshitz, Yad. Fiz. \textbf{35}, 1316 (1982);
Proc. CORINNE 90, Nantes, France, 1990 (ed. D. Ardouin, World Sci., 1990) p. 42.

\bibitem{fsi} J. Adams \textit{et. al.} (STAR), Phys. Rev. C, \textbf{74}, 064906, (2006).

\bibitem{BW} P. J. Siemens and J. O. Rasmussen, Phys. Rev. Lett. \textbf{42}, 880 (1979);
             E. Schnedermann, J. Sollfrank, and U. Heinz, PRC \textbf{48}, 2462 (1993);
             A. Kisiel, Braz. J. Phys. \textbf{37}, 917 (2007).

\bibitem{HKM} Yu.M. Sinyukov, S.V. Akkelin, and Y. Hama, Phys. Rev. Lett. \textbf{89}, 052301 (2002).

\bibitem{Boltz} S.V. Akkelin, Y. Hama, Iu.A. Karpenko, Yu.M. Sinyukov. Phys. Rev. C \textbf{78}, 034906, (2008).

\bibitem{Kaon} Iu.A. Karpenko, Yu.M. Sinyukov. Phys. Rev. C {\bf 81} 054903, (2010).

\bibitem{Uniform} Iu.A. Karpenko, Yu.M. Sinyukov, K. Werner. Phys. Rev. C \textbf{87}, 024914, (2013), \\ arXiv:1204.5351v2 [nucl-th].

\bibitem{Emiss} S.V. Akkelin, Yu.M. Sinyukov, Phys. Rev. C \textbf{73}, 034908 (2006).

\bibitem{Groot} S. R. de Groot, W. A. van Leeuwen, and Ch. G. van Weert, Relativistic Kinetic Theory (North-Holland, Amsterdam, 1980).

\bibitem{Fermi} E. Fermi, Z. Phys. \textbf{88}, 161 (1934); translated in F.L. Wilson, Am. J. Phys. \textbf{36}, 1150 (1968).

\bibitem{Koonin} S. E. Koonin, Phys. Lett. B, \textbf{70}, 43 (1977).

\bibitem{Gyulassy} M. Gyulassy, S.K. Kauffmann and L.W. Wilson, Phys. Rev. C \textbf{20}, 2267 (1979).

\bibitem{Ledn1} R. Lednick\'y, V.V. Lyuboshitz and V.L. Lyuboshitz, Phys. Atom. Nucl. 61, 2050 (1998); \\
R. Lednick\'y, J. Phys. G: Nucl. Part. Phys. 35, 125109 (2008); 

\bibitem{Ledn2} R. Lednick\'y, Phys. Part. Nuclei 40, 307 (2009).

\bibitem{Ledn3} R. Lednick\'y, Phys. Part. Nuclei Lett. 8, 965 (2011).

\bibitem{harms} D.A. Brown, P. Danielewicz, A. Enokizono, M. Heffner, R. Soltz, S. Pratt, Phys. Rev. C, \textbf{72},  054902 (2005).

\bibitem{Offset} R. Lednick\'y, V.L. Lyuboshitz, B. Erazmus, and D. Nouais, Phys. Lett. B \textbf{373}, 30 (1996).

\bibitem{DanPratt} P. Danielewicz and S. Pratt, Phys. Rev. C \textbf{75}, 034907 (2007).

\bibitem{Chaj} Z. Chajecki and M. Lisa, Braz. J. Phys. \textbf{37}, 1057 (2007).

\bibitem{Cooper} F. Cooper and G. Frye, Phys. Rev. D \textbf{10}, 186 (1974).

\bibitem{glissando} W. Broniowski, M. Rybczynsky, P. Bozek, arXiv:0710.5731v3

\bibitem{Phenix2} S.S. Adler \textit{et al.} PHENIX Collaboration, Phys. Rev. Lett. 98, 132301 (2007). 

\bibitem{AnDiff} M. Csanad, T. Cs\"org\H{o}, M. Nagy, Braz. J. Phys. \textbf{37}, 1002 (2007).

\bibitem{RoRs1} M. S. Borysova, Yu. M. Sinyukov, S. V. Akkelin, B. Erazmus, and Iu. A. Karpenko,
			   Phys. Rev. C \textbf{73}	, 024903 (2006).

\bibitem{RoRs2} Iu.A. Karpenko, Yu.M. Sinyukov, Phys. Lett. B \textbf{688} 50 (2010).

\bibitem{Pratt2} S. Pratt, Phys. Rev. Lett. \textbf{53}, 1219, 1984.

\bibitem{Rischke} Dirk H. Rischke \textit{et al.}, Nucl. Phys. A \textbf{608}, 479, 1996.

\end{thebibliography}
\end{document}